\documentclass[12pt]{iopart}

\usepackage{graphicx}
\usepackage{epsfig}

\begin{document}

\title[VBS on Arbitrary Graph]{Entanglement of Valence-Bond-Solid on an Arbitrary Graph}

\author{Ying Xu$^{1}$ and Vladimir E Korepin$^{2}$}

\address{C.N. Yang Institute for Theoretical Physics \\
 State University of New York at Stony Brook, Stony Brook, NY 11794-3840, USA\\}
\ead{$^{1}$\mailto{yixu@ic.sunysb.edu}, $^{2}$\mailto{korepin@insti.physics.sunysb.edu}}

\begin{abstract}
The Affleck-Kennedy-Lieb-Tasaki (AKLT) spin interacting model can be defined on an arbitrary graph. We explain the construction of the AKLT Hamiltonian. Given certain conditions, the ground state is unique and known as the Valence-Bond-Solid (VBS) state. It can be used in measurement-based quantum computation as a resource state instead of the cluster state. We study the VBS ground state on an arbitrary connected graph. The graph is cut into two disconnected parts: the block and the environment. We study the entanglement between these two parts and prove that many eigenvalues of the density matrix of the block are zero. We describe a subspace of eigenvectors of the density matrix corresponding to non-zero eigenvalues. The subspace is the degenerate ground states of some Hamiltonian which we call the block Hamiltonian.
\end{abstract}

\vspace{2pc}
\submitto{\JPA}
\maketitle

\section{Introduction}
\label{secintroduction}

The fields of statistical physics, condensed matter physics and quantum information theory share a common interest in the study of interacting quantum many body systems. The concept of entanglement in quantum mechanics has significant importance in all these areas. Much of the current effort is devoted to the description and quantification of the entanglement contained in strongly correlated quantum states. Quantum entanglement is a fundamental measure of how much quantum effects we can observe and use to control one quantum system by another, and it is the primary resource in quantum computation and quantum information processing \cite{BD}, \cite{L}. Entanglement properties play an important role in condensed matter physics, such as phase transitions \cite{ON}, \cite{OAFF} and macroscopic properties of solids \cite{GRAC}. Extensive research has been undertaken to investigate quantum entanglement for spin chains, correlated electrons, interacting bosons as well as other models, see \cite{AFOV}, \cite{AEPW}, \cite{BDV}, \cite{CDR}, \cite{CZWZ}, \cite{FL}, \cite{GLL}, \cite{GDLL}, \cite{HZS}, \cite{H2}, \cite{HLW}, \cite{JK}, \cite{KM}, \cite{KP}, \cite{K}, \cite{LO}, \cite{LORV}, \cite{LRV}, \cite{LW}, \cite{O1}, \cite{OL}, \cite{PP}, \cite{PHE}, \cite{PS}, \cite{RH}, \cite{VMC}, \cite{VPC}, \cite{VDB}, \cite{WK}, \cite{WKRL}, \cite{ZR} for reviews and references. Characteristic functions of quantum entanglement, such as von Neumann entropy and Renyi entropy, were obtained and discussed through studying reduced density matrices of subsystems \cite{FKR}, \cite{FIJK}, \cite{FIK}, \cite{IJK}, \cite{KHH}. An area law for the von Neumann entropy in harmonic lattice systems has been extensively studied \cite{CEP}, \cite{CEPD}, \cite{Has}, \cite{PEDC}. 

Much insight in understanding entanglement of quantum systems has been obtained by studying exactly solvable models in statistical mechanics. In 1987, I. Affleck, T. Kennedy, E. H. Lieb and H. Tasaki proposed a spin interacting model known as the AKLT model \cite{AKLT1}, \cite{AKLT2}. The model consists of spins on a lattice and the Hamiltonian describes interactions between nearest neighbors. The Hamiltonian density is a linear combination of projectors. The model is similar to the Heisenberg anti-ferromagnet with a gap. The authors (AKLT) of \cite{AKLT1}, \cite{AKLT2} found the exact ground state, which has an exponentially-decaying correlation function and a finite energy gap. This model has been attracting enormous research interests since then \cite{CCK}, \cite{FK}, \cite{FKRHB}, \cite{KHK}, \cite{IT}. It can be defined and solved in higher dimensional and arbitrary lattices \cite{AKLT2}, \cite{FKR2}, \cite{KLT}, \cite{RB} and generalizable to the inhomogeneous (non-translational invariant) case (spins at different lattice sites may take different values) and an arbitrary graph \cite{KK}. Given certain conditions (as to be described later), the ground state has proven to be unique \cite{AAH}, \cite{KK}. It is known as the Valence-Bond-Solid (VBS) state. The Schwinger boson representation of the VBS state (see (\ref{generalizedvbs})) relates to the Laughlin ansatz of the fractional quantum Hall effect \cite{AAH}, \cite{HR}, \cite{HZS}, \cite{ILO}, \cite{LA}. The Laughlin wave function of the fractional quantum Hall effect is the VBS state on the complete graph \cite{CTZ}, \cite{HR0}. The VBS state illustrates ground state properties of anti-ferromagnetic integer-spin chains with a Haldane gap \cite{H}. The theory of VBS state was essentially developed by B. Nachtergaele and others \cite{FNW0}, \cite{FNW1}, \cite{FNW}, \cite{FNW2}, \cite{N1}, \cite{N2}. The Entanglement of formation in VBS state was estimated in \cite{MN}. Brennen and Miyake showed that the VBS state can be used as a resource state in measurement-based quantum computing instead of the cluster state \cite{BM}. It was proved in \cite{VC} that VBS state allows universal quantum computation and an implementation of the AKLT Hamiltonian in optical lattices \cite{GMC} has also been proposed. 

We shall consider a part of the system, i.e. a block of spins. It is described by the density matrix of the block, which we call \textit{the density matrix} later for short. The density matrix  has been studied extensively in \cite{FKR}, \cite{FM}, \cite{KHH}, \cite{KK}, \cite{VMC}. It contains information of all correlation functions \cite{AAH}, \cite{JK}, \cite{KHH}, \cite{XKHK}. Furthermore, the entanglement properties of the VBS ground state has been studied by means of the density matrix as in \cite{FK}, \cite{FKR}, \cite{FKR2}, \cite{H2}, \cite{KHH}, \cite{VMC}. The von Neumann entropy of the subsystem density matrix is a measure of entanglement of the VBS state. The Renyi entropy is another measure of the entanglement. The entanglement entropy was obtained in \cite{FKR}, \cite{FM}, \cite{KHH}, \cite{XKHK}.

The structure of the density matrix is important. For a $1$-dimensional AKLT spin chain the density matrix has a lot of zero eigenvalues \cite{XKHK}, \cite{XKHK2}. The eigenvectors with non-zero eigenvalues are the degenerate ground states of some Hamiltonian, which we shall call the \textit{block} Hamiltonian (see (\ref{subsystemh})). In the limit of large block, the density matrix is proportional to a projector on the degenerate ground states of the block Hamiltonian. These states are the only eigenstates of the density matrix with non-zero eigenvalues which contribute to the entropy. Also, eigenstates of the density matrix are useful in quantum computing because of quantum measurements. It was conjectured in \cite{XKHK} that eigenvectors of the density matrix with non-zero eigenvalues always form degenerate ground states of some Hamiltonian (the block Hamiltonian), which is generalizable to an arbitrary graph. In this paper we shall give a general proof of this statement.

The paper is divided into five parts:
\begin{enumerate}
	\item We define the \textit{basic} AKLT model on an arbitrary connected graph and construct the unique VBS ground state using symmetrization and anti-symmetrization of states. A graphical illustration is included. (Section \ref{secbasic})
	\item We introduce the \textit{generalized} (inhomogeneous) AKLT model and give the condition of the uniqueness of the ground state. The VBS ground state is constructed using the Schwinger boson representations. Within this formulation, the relation between the VBS state and the Laughlin states of the factional quantum Hall effect becomes obvious \cite{AAH}, \cite{CTZ}, \cite{HR0}, \cite{HR}, \cite{HZS}, \cite{LA}. (Section \ref{secgeneralized})
	\item In order to study entanglement of the VBS state, we cut the graph into two disconnected parts: the \textit{block} and the \textit{environment}. We define the \textit{block} Hamiltonian, and show that its ground state is degenerate. (Section \ref{secsubsystem})
	\item The density matrix of the block is proved to have a lot of zero eigenvalues. The eigenvectors with non-zero eigenvalues form degenerate ground states of the block Hamiltonian. (Section \ref{secmatrix})
	\item Examples of the density matrix are given explicitly as special cases of the general result. We also formulate some open problems. (Section \ref{secexamples})
\end{enumerate}

In Sections \ref{secbasic} and \ref{secgeneralized} we follow the paper of A. N. Kirillov and V. E. Korepin written in 1990, see \cite{KK}.

\section{The Basic AKLT Model}
\label{secbasic}

\begin{figure}
	\centering
	\includegraphics[width=3in]{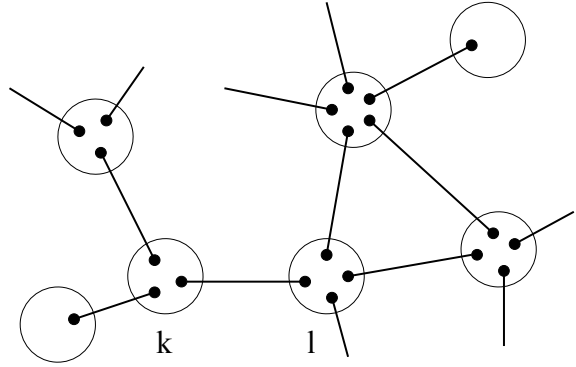}
	\caption{Example of a part of the graph including vertex $k$ with $z_{k}=3$ and vertex $l$ with $z_{l}=4$. Black dots \fullcircle represent spin-$\frac{1}{2}$ states, which are enclosed by large circles \opencircle representing vertices and symmetrization of the product of spin-$\frac{1}{2}$'s at each vertex. Solid lines \full \ represent edges which anti-symmetrize the pair of connected spin-$\frac{1}{2}$'s.}
	\label{figure1}
\end{figure}

We start with the definition of the \textit{basic} AKLT model on a connected graph. A \textit{graph} consists of two types of elements, namely \textit{vertices} and \textit{edges}. Every edge \textit{connects} two vertices. As in Figure (\ref{figure1}), a vertex is drawn as a (large) circle \opencircle and an edge is drawn as a solid line \full\ connecting two vertices. For every pair of vertices in the \textit{connected} graph, there is a \textit{walk} \footnote{A \textit{walk} is an alternating sequence of vertices and edges, beginning and ending with a vertex, in which each vertex is incident to the two edges that precede and follow it in the sequence, and the vertices that precede and follow an edge are the endvertices of that edge.} from one to the other. Vertices can also be called \textit{sites} and edges sometimes called \textit{links} or \textit{bonds}. In the case of a disconnected graph, the Hamiltonian (\ref{basich}) is a direct sum with respect to connected components and the ground state is a direct product. We shall start with a connected graph. We shall also assume that the graph consists of more than one vertices, otherwise there would be no interaction at all. Let's introduce notations. By $\bi{S}_{l}$ we shall denote the spin operator located at vertex $l$ with spin value $S_{l}$. In the \textit{basic} model we require that $S_{l}=\frac{1}{2}z_{l}$, where $z_{l}$ is the number of incident edges (connected to vertex $l$), also known as the valence or coordination number (the number of nearest neighbors of the vertex $l$). The relation between the spin value and coordination number must be true for any vertex $l$, including boundaries. This will guarantee uniqueness of the ground state. The Hamiltonian describes interactions between nearest neighbors:
\begin{eqnarray}
	H=\sum_{\langle kl\rangle}H(k,l). \label{basich}
\end{eqnarray}
Here $H(k,l)$ describes the interaction between spins at vertices $k$ and $l$ connected by an edge, and we sum over all edges $\langle kl\rangle$. The Hamiltonian density is $H(k,l)$. To write down an explicit form of $H(k,l)$, we define a projector $\pi_{J}(k,l)$:
\begin{eqnarray}
	\pi_{J}(k,l)=\prod^{j\neq J}_{|S_{k}-S_{l}|\leq j\leq S_{k}+S_{l}}\frac{(\bi{S}_{k}+\bi{S}_{l})^{2}-j(j+1)}{J(J+1)-j(j+1)}. \label{projectorpi}
\end{eqnarray}
Operator $\pi_{J}(k,l)$ projects the edge spin $\bi{J}_{kl}\equiv\bi{S}_{k}+\bi{S}_{l}$ on the subspace with fixed total spin value $J$ and $|S_{k}-S_{l}|\leq J\leq S_{k}+S_{l}$. Note that we could expand $(\bi{S}_{k}+\bi{S}_{l})^{2}=2\bi{S}_{k}\cdot\bi{S}_{l}+S_{k}(S_{k}+1)+S_{l}(S_{l}+1)$. So that projector $\pi_{J}(k,l)$ in (\ref{projectorpi}) is a polynomial in the scalar product $(\bi{S}_{k}\cdot\bi{S}_{l})$ of degree $2S_{min}$, where $S_{min}\equiv Min\{S_{k}, S_{l}\}$ is the minimum of the two spin values of the same edge. For example with $S_{k}=S_{l}=1$, we may have a quadratic polynomial:
\begin{eqnarray}
	\pi_{2}(k,l)=\frac{1}{6}(\bi{S}_{k}\cdot\bi{S}_{l})^{2}+\frac{1}{2}(\bi{S}_{k}\cdot\bi{S}_{l})+\frac{1}{3}. \label{examplepi2}
\end{eqnarray}
In the \textit{basic} model we define the Hamiltonian density $H(k,l)$ as
\begin{eqnarray}
	H(k,l)=A(k,l)\ \pi_{S_{k}+S_{l}}(k,l), \qquad H(k,l)\geq 0 \label{basichd}
\end{eqnarray}
with $A(k,l)$ an arbitrary positive real coefficient (it may depend on the edge $\langle kl\rangle$). So that the Hamiltonian density for each edge is proportional to the projector on the subspace with the highest possible edge spin value $(S_{k}+S_{l})$. The physical meaning is that interacting spins do not form the highest possible edge spin (this will increase the energy) in the ground state. The Hamiltonian in (\ref{basich}) is a linear combination of projectors with positive coefficients, which shows that $H$ is semi-positive definite.

The Hamiltonian (\ref{basich}) with condition
\begin{eqnarray}
	S_{l}=\frac{1}{2}z_{l} \label{basiccond}
\end{eqnarray}
has a unique ground state \cite{AKLT1}, \cite{AKLT2}, \cite{AAH}, \cite{KK} known as the Valence-Bond-Solid (VBS) state. It can be constructed as follows.
Each vertex $l$ has $z_{l}$ spin-$\frac{1}{2}$'s. We associate each spin-$\frac{1}{2}$ with an incident edge. In such a way each edge has two spin-$\frac{1}{2}$'s at its ends. We anti-symmetrize the wave function of these two spin-$\frac{1}{2}$'s. So that anti-symmetrization is done along each edge. We also symmetrize the product of spin-$\frac{1}{2}$'s at each vertex (each large circle). Let's write down the VBS ground state algebraically. We label the particular dot from vertex $l$ connected with some dot from vertex $k$ by $l_{k}$ (correspondingly, that dot from vertex $k$ is labeled by $k_{l}$). In this way we have specified a unique prescription of labels with dots. Then the anti-symmetrization results in the singlet state
\begin{eqnarray}
	|\Phi\rangle_{kl}=\frac{1}{\sqrt{2}}\left(|\uparrow\rangle_{l_{k}}|\downarrow\rangle_{k_{l}}-|\downarrow\rangle_{l_{k}}|\uparrow\rangle_{k_{l}}\right). \label{antisymm}
\end{eqnarray}
The direct product of all these $|\Phi\rangle$ singlet states corresponds to all edges in our graph:
\begin{eqnarray}
	\prod_{\langle kl\rangle}|\Phi\rangle_{kl}. \label{prodsing}
\end{eqnarray}
We still have to complete the symmetrization (circles) at each vertex. We denote the symmetrization operator of $z_{l}$ dots in vertex $l$ by $\mathbf{P}(l)$, then the symmetrization at each vertex is carried out by taking the product $\prod_{l}\mathbf{P}(l)$ of all vertices. Finally, the unique VBS ground state can be written as
\begin{eqnarray}
	|\mbox{VBS}\rangle=\prod_{l}\mathbf{P}(l)\prod_{\langle kl\rangle}|\Phi\rangle_{kl}. \label{basicvbs}
\end{eqnarray}
Here the first product runs over all vertices and the second over all edges. If the coordination number $z_{l}$ is a constant over all vertices in the graph except for boundaries, then we would have the same spin value at each bulk vertex. In that case the \textit{basic} model is also referred to as the \textit{homogeneous} model.

\section{The Generalized AKLT Model}
\label{secgeneralized}

In the \textit{generalized} AKLT model, relation (\ref{basiccond}) is generalized. We associate a positive integer $M_{kl}$ ($M_{kl}\equiv M_{lk}$) to each edge $\langle kl\rangle$ of the graph. We shall call $M_{kl}$ multiplicity numbers. The Hamiltonian describes interactions between nearest neighbors (vertices connected by an edge):
\begin{eqnarray}
	H=\sum_{\langle kl\rangle}H(k,l). \label{generalizedh}
\end{eqnarray}
However, the Hamiltonian density is no longer proportional to a single projector in general. It is a linear combination of projectors
\begin{eqnarray}
	H(k,l)=\sum^{S_{k}+S_{l}}_{J=S_{k}+S_{l}-M_{kl}+1}A_{J}(k,l)\ \pi_{J}(k,l), \qquad H(k,l)\geq 0. \label{generalizedhd}
\end{eqnarray}
Projector $\pi_{J}(k,l)$ is given by (\ref{projectorpi}), and $A_{J}(k,l)$'s are arbitrary positive coefficients. So that $H(k,l)$ projects the edge spin on the subspace with spin value $J$ greater than $S_{k}+S_{l}-M_{kl}$. Physically formation of edge spin higher than $S_{k}+S_{l}-M_{kl}$ would increase the energy. Cappelli, Trugenberger and Zemba showed that the Hamiltonian for the fractional quantum Hall effect also can be written as a linear combination of projectors \cite{CTZ}.

The condition of uniqueness of the ground state was introduced in \cite{KK}:
\begin{eqnarray}
	2S_{l}=\sum_{k}M_{kl}, \qquad \forall \ l. \label{condition1}
\end{eqnarray}
Here $S_{l}$ is the spin value at vertex $l$ and we sum over all edges incident to vertex $l$ (connected to vertex $l$). The Hamiltonian (\ref{generalizedh}) has a unique ground state if (\ref{condition1}) is valid. The relation $2S_{l}=z_{l}$ for the \textit{basic} model is a special case when $M_{kl}=1$. The condition (\ref{condition1}) can be put into an invariant form. Let's define a column vector $\mathbf{ S}$, the $l^{th}$ component of which is associated with vertex $l$ of the graph and equal to $S_{l}$. The number of components is equal to the number of vertices $N$. Next, we define another column vector $\mathbf{M}$ with its dimension equal to the number of edges $M$ in the graph. The $k^{th}$ and $l^{th}$ components of this vector are associated with edge $\langle kl\rangle$ and both equal to $M_{kl}$. The most important geometrical characteristic of the graph is the vertex-edge incidence matrix $\hat{\mathbf{I}}$ (see \cite{Har}). This is a rectangular matrix with $N$ rows and $M$ columns. Each row is associated with the vertex and each column is associated with the edge. If the vertex belongs to the edge the corresponding matrix element is equal to one, otherwise zero. Then the condition (\ref{condition1}) of uniqueness can be re-written as
\begin{eqnarray}
	2\ \mathbf{S}=\hat{\mathbf{I}}\cdot\mathbf{M}. \label{condition2} 
\end{eqnarray}
For more details we refer to \cite{KK}.

Under condition (\ref{condition1}) or (\ref{condition2}), the unique ground state of Hamiltonian (\ref{generalizedh}) is referred to as the generalized VBS state. It is constructed by introducing the Schwinger boson representation \cite{AAH}, \cite{FM}, \cite{KK}, \cite{KHH}, \cite{XKHK}, \cite{XKHK2}. We define a pair of independent canonical bosonic operators $a_{l}$ and $b_{l}$ for each vertex $l$:
\begin{eqnarray}
	[\ a_{k},\ a^{\dagger}_{l}\ ]=[\ b_{k},\ b^{\dagger}_{l}\ ]=\delta_{kl} \label{commutator}
\end{eqnarray}
with all other commutators vanishing:
\begin{eqnarray}
	[\ a_{k},\ a_{l}\ ]=[\ b_{k},\ b_{l}\ ]=[\ a_{k},\ b_{l}\ ]=[\ a_{k},\ b^{\dagger}_{l}\ ]=0, \qquad \forall\ k, l. \label{commvani}
\end{eqnarray}
Spin operators are represented as 
\begin{eqnarray}
	S^{+}_{l}=a^{\dagger}_{l}b_{l}, \qquad S^{-}_{l}=b^{\dagger}_{l}a_{l}, \qquad S^{z}_{l}=\frac{1}{2}(a^{\dagger}_{l}a_{l}-b^{\dagger}_{l}b_{l}). \label{spinrep}
\end{eqnarray}
To reproduce the dimension of the spin-$S_{l}$ Hilbert space at vertex $l$, a constraint on the total boson occupation number is required:
\begin{eqnarray}
	\frac{1}{2}(a^{\dagger}_{l}a_{l}+b^{\dagger}_{l}b_{l})=S_{l}. \label{constraint}
\end{eqnarray}
As a result, the VBS ground state in the Schwinger representation is given by
\begin{eqnarray}
	|\mbox{VBS}\rangle=
 \prod_{\langle kl\rangle}
\left(a^{\dagger}_{k}b^{\dagger}_{l}-b^{\dagger}_{k}a^{\dagger}_{l}\right)^{M_{kl}}|\mbox{vac}\rangle, \label{generalizedvbs}
\end{eqnarray}
This representation shows that for a full graph (each vertex is connected to every other vertex) the VBS state coincides with the Laughlin wave function \cite{AAH}, \cite{CTZ}, \cite{HR0}, \cite{HR}, \cite{HZS}. In (\ref{generalizedvbs}) the product runs over all edges and the \textit{vacuum} $\left|\mbox{vac}\right\rangle$ is annihilated by any of the annihilation operators:
\begin{eqnarray}
	a_{l}\left|\mbox{vac}\right\rangle=b_{l}\left|\mbox{vac}\right\rangle=0, \qquad \forall \ l. \label{vacuum}
\end{eqnarray}
Note that $[\ a^{\dagger}_{k},\ b^{\dagger}_{l}\ ]=0$, $\forall\ k,l$. If we replace $a^{\dagger}$'s and $b^{\dagger}$'s by complex numbers (coordinates of electrons), then (\ref{generalizedvbs}) turns into the Laughlin wave function of the fractional quantum Hall effect \cite{AAH}, \cite{HR}. To prove that (\ref{generalizedvbs}) is the ground state we need only to verify for any vertex $l$ and edge $\langle kl\rangle$: (i) the total power of $a^{\dagger}_{l}$ and $b^{\dagger}_{l}$ is $2S_{l}$, so that we have spin-$S_{l}$ at vertex $l$; (ii) $-\frac{1}{2}(\sum_{l^{\prime}\neq l}M_{l^{\prime}k}+\sum_{k^{\prime}\neq k}M_{k^{\prime}l})\leq J^{z}_{kl}\equiv S^{z}_{k}+S^{z}_{l}\leq \frac{1}{2}(\sum_{l^{\prime}\neq l}M_{l^{\prime}k}+\sum_{k^{\prime}\neq k}M_{k^{\prime}l})$ by a binomial expansion, so that the maximum value of the edge spin $J_{kl}$ is $\frac{1}{2}(\sum_{l^{\prime}\neq l}M_{l^{\prime}k}+\sum_{k^{\prime}\neq k}M_{k^{\prime}l})=S_{k}+S_{l}-M_{kl}$ (from $SU(2)$ invariance, see \cite{AAH}). Therefore, the state $|\mbox{VBS}\rangle$ defined in (\ref{generalizedvbs}) has spin-$S_{l}$ at vertex $l$ and no projection onto the $J_{kl}>S_{k}+S_{l}-M_{kl}$ subspace for any edge. The introduction of Schwinger bosons can be used to construct a spin coherent state basis in which spins at each vertex behave as classical unit vectors. The spin coherent state basis is given in \ref{secapp}.

\section{The Entanglement between Block and Environment}
\label{secsubsystem}

The VBS state (see (\ref{basicvbs}) and (\ref{generalizedvbs})) has non-trivial entanglement properties. The density matrix of the VBS state is a projector
\begin{eqnarray}
	\rho=\frac{|\mbox{VBS}\rangle\langle \mbox{VBS}|}{\langle \mbox{VBS}|\mbox{VBS}\rangle}. \label{purematrix}
\end{eqnarray}
Let us cut the original graph into two subgraphs $B$ and $E$, that is, we cut through some number of edges such that the resulting graph $B\cup E$ is disconnected (no edge between $B$ and $E$). We may call one of them, say $B$, the \textit{block}, and the other one $E$ the \textit{environment}. The distinction is arbitrary and the two subsystems are equivalent in measuring entanglement. 

Let's focus on the block (subsystem $B$). It is described by the density matrix $\rho_{b}$ of the block (obtained by tracing out degrees of freedom of the environment $E$ from the density matrix $\rho$ (\ref{purematrix})):
\begin{eqnarray}
	\rho_{b}=\Tr_{e}[\ \rho\ ]. \label{rhoa}
\end{eqnarray}
In (\ref{rhoa}) and below we use subscript $b$ for \textit{block} and $e$ for \textit{environment}. The density matrix $\rho_{b}$ contains all correlation functions in the VBS ground state as matrix entries \cite{AAH}, \cite{JK}, \cite{KHH}, \cite{XKHK}. The entanglement can be measured by the von Neumann entropy
\begin{eqnarray}
	S_{v. N}=-\Tr_{b}[\ \rho_{b}\ln\rho_{b}\ ]=-\sum_{\lambda\neq 0}\lambda\ln\lambda \label{vnentropy}
\end{eqnarray}
or the Renyi entropy
\begin{eqnarray}
	S_{R}(\alpha)=\frac{1}{1-\alpha}\ln\left\{\Tr_{b}[\ \rho^{\alpha}_{b}\ ]\right\}=\frac{1}{1-\alpha}\ln\left(\sum_{\lambda\neq 0}\lambda^{\alpha}\right), \qquad \alpha>0.
\end{eqnarray}
Here $\lambda$'s are (non-zero) eigenvalues of density matrix $\rho_{b}$ and $\alpha$ is an arbitrary parameter. It was shown by using the Schmidt decomposition \cite{NC} that non-zero eigenvalues of the density matrix of subsystem $B$ (block) is equal to those of the density matrix of subsystem $E$ (environment). So the two subsystems are equivalent in measuring entanglement in terms of entanglement entropies, i.e. $S_{v. N}[B]=S_{v. N}[E]$ and $S_{R}[B]=S_{R}[E]$. This fact has been used in obtaining entanglement entropies of $1$-dimensional VBS states as in \cite{FKR}, \cite{KHH}. We shall show that the spectrum of the density matrix $\rho_{b}$ contains a lot of zero eigenvalues. In order to understand this and give the subsystem (block) a more complete description, we first introduce the Hamiltonian of the subsystem (called the \textit{block} Hamiltonian).

\begin{figure}
	\centering
	\includegraphics[width=2.5in]{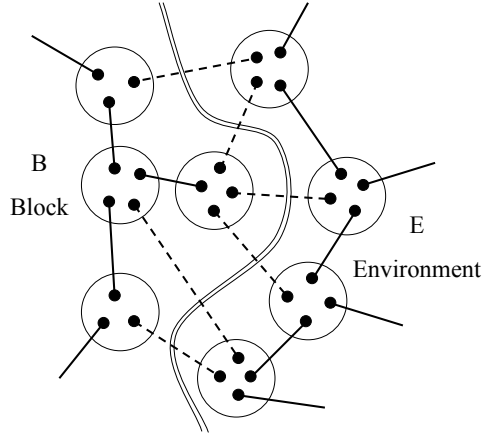}
	\caption{Example of the cutting for the basic model. The curved double line represents the boundary between the two subgraphs. We have the block $B$ on the left and the environment $E$ on the right. Solid lines	\full\ represent edges while dashed lines \dashed\ represent cut edges. Each dashed line connects two dots. All vertices in the figure belong to the boundary of $B$ or $E$ because of the presence of one or more cut incident edges (dashed lines).}
	\label{figure2}
\end{figure}

The \textit{block} Hamiltonian $H_{b}$ is the sum of Hamiltonian densities $H(k,l)$ with both $k\in B$ and $l\in B$, i.e. nearest neighbor interactions (edge terms) within the block $B$:
\begin{eqnarray}
	H_{b}=\sum_{\langle kl\rangle\in B}H(k,l),\qquad k\in B, \quad l\in B. \label{subsystemh}
\end{eqnarray}
Here $H(k,l)$ is given in (\ref{basichd}) for the basic model and (\ref{generalizedhd}) for the generalized model, for $k$ and $l$ connected by an edge. In (\ref{subsystemh}) no cut edges are present (boundary edges between subgraphs $B$ and $E$ removed). This Hamiltonian has degenerate ground states because uniqueness conditions (\ref{basiccond}) and (\ref{condition1}) are not valid. Let us discuss the degeneracy of ground states of (\ref{subsystemh}). Let's denote by $L$ the number of vertices on the boundary of the block $B$. The boundary consists of those vertices with one or more cut incident edge, see Figure (\ref{figure2}).
The degeneracy $deg.$ of ground states of $H_{b}$ is given by the Katsura's formula \cite{Kcomm}
\begin{eqnarray}
	deg.=\prod_{l\in \partial B}\left[\left(\sum_{k \in \partial E}M_{kl}\right)+1\right], \qquad \langle kl\rangle\in\{\mbox{cut edges}\}. \label{generalizeddeg}
\end{eqnarray}
Here $\partial B$ denotes vertices on the boundary of the block $B$ and $\partial E$ are vertices on the boundary of the environment $E$. In (\ref{generalizeddeg}) we have $L$ terms in the product. Formula (\ref{generalizeddeg}) is valid for both the basic and the generalized model. For the basic model all $M_{kl}=1$. A proof of formula (\ref{generalizeddeg}) is given in \ref{secapp2}. The subspace spanned by the degenerate ground states is called the \textit{ground space}, with the dimension given by $deg.$ in (\ref{generalizeddeg}). We emphasize at this point that the block $B$ should contain more than one vertices, otherwise the block Hamiltonian vanishes $H_{b}=0$ and the whole Hilbert space become the \textit{ground space}. We discuss the density matrix for a single vertex block at the end of next section. It was shown for $1$-dimensional models in \cite{XKHK}, \cite{XKHK2} that the spectrum of density matrix $\rho_{b}$ is closely related to the block Hamiltonian. The density matrix is a projector onto the ground space multiplied by another matrix. We shall prove the statement for an arbitrary graph in the next section.

\section{The Density Matrix}
\label{secmatrix}

Let us denote by $N_{b}$ the number of vertices in the block $B$. Then the dimension $dim.$ of the Hilbert space of the block $B$ is equal to $\prod_{l}(2S_{l}+1)$ with $l\in B$, which is also the dimension of the density matrix $\rho_{b}$. The value is
\begin{eqnarray}
	dim.=\prod_{l\in B}\left[z_{l}+1\right], \label{basicdima}
\end{eqnarray}
for the basic model and
\begin{eqnarray}
	dim.=\prod_{l\in B}\left[\left(\sum_{k\in (B\cup\partial E)}M_{kl}\right)+1\right], \label{generalizeddima}
\end{eqnarray}
for the generalized model. In both expressions (\ref{basicdima}) and (\ref{generalizeddima}) we have $N_{b}$ factors in the product. The density matrix $\rho_{b}$ would have $dim.$ number of eigenvalues. However, most of the eigenvalues are vanishing and $\rho_{b}$ is a projector onto a much smaller subspace multiplied by another matrix. To prove the statement, we define a \textit{support} to be the subspace of the Hilbert space of the block $B$ with non-zero eigenvalues, i.e. it is spanned by eigenstates of $\rho_{b}$ with non-zero eigenvalues. The dimension of the support is denoted by $D$. We have the following theorem on the structure of the density matrix $\rho_{b}$ (Assuming that the block have more than one vertices, i.e. $N_{b}\geq2$, so that $H_{b}$ is not equal to zero identically):

\textbf{Theorem}
The \textit{support} of $\rho_{b}$ (\ref{rhoa}) is a subspace of the \textit{ground space} of the block Hamiltonian $H_{b}$ (\ref{subsystemh}).

To prove the theorem, we recall that $H=\sum_{\langle kl\rangle\in B}H(k,l)$ and each $H(k,l)$ is a sum of projectors (\ref{generalizedhd}). We have $H(k,l)\geq 0$. Then the construction of the VBS ground state (\ref{basicvbs}) and (\ref{generalizedvbs}) guarantees that there is no projection onto the subspace with higher edge spins for \textit{any} edge. Therefore,
\begin{eqnarray}
	H(k,l)|\mbox{VBS}\rangle=0, \qquad \forall\ \langle kl\rangle. \label{noprojection}
\end{eqnarray}
In particular, this is true for edges inside the block $B$, i.e. $k\in B$ and $l\in B$. Now, from the definition of $\rho_{b}$ in ($\ref{rhoa}$), we have
\begin{eqnarray}
	H(k,l)\rho_{b}&=&H(k,l)\Tr_{e}[\ \rho\ ]
\nonumber \\
&=&\frac{H(k,l)\Tr_{e}[\ |\mbox{VBS}\rangle\langle \mbox{VBS}|\ ]}{\langle \mbox{VBS}|\mbox{VBS}\rangle}
\nonumber \\
&=&\frac{\Tr_{e}[\ H(k,l)|\mbox{VBS}\rangle\langle \mbox{VBS}|\ ]}{\langle \mbox{VBS}|\mbox{VBS}\rangle}=0, \qquad k\in B, \quad l\in B. \label{proof}
\end{eqnarray}
In the last step of (\ref{proof}) we have used (\ref{noprojection}) and the fact that edge $\langle kl\rangle$ lies completely inside block $B$ so that $H(k,l)$ commutes with tracing in the environment $E$. Equation (\ref{proof}) is true for any edge in $B$, so that
\begin{eqnarray}
	H_{b}\rho_{b}=\sum_{\langle kl\rangle\in B}H(k,l)\rho_{b}=0, \qquad k\in B, \quad l\in B. \label{harhoa}
\end{eqnarray}
If we diagonalize the density matrix $\rho_{b}$
\begin{eqnarray}
	\rho_{b}=\sum_{\lambda\neq 0}\lambda|\lambda\rangle\langle \lambda|, \label{diag}
\end{eqnarray}
where $|\lambda\rangle$ is the eigenstate corresponding to eigenvalue $\lambda$. Then (\ref{harhoa}) can be re-written as
\begin{eqnarray}
	H_{b}\sum_{\lambda\neq 0}\lambda|\lambda\rangle\langle \lambda|=\sum_{\lambda\neq 0}\lambda H_{b}|\lambda\rangle\langle \lambda|=0, \label{eigenstate}
\end{eqnarray}
Note that $\{|\lambda\rangle\}$ is a linearly independent set. Therefore the solution of (\ref{eigenstate}) means that
\begin{eqnarray}
	H_{b}|\lambda\rangle=0, \qquad \lambda\neq 0. \label{solution}
\end{eqnarray}
Expression (\ref{solution}) states that any eigenstate of $\rho_{b}$ (with non-zero eigenvalue) is a ground state of $H_{b}$. As a result, we have proved the \textbf{Theorem} that the \textit{support} of $\rho_{b}$ is a subspace of the \textit{ground space} of $H_{b}$, so that $D\leq deg.$ The density matrix takes the form of a projector multiplied by another matrix and the projector projects on the \textit{ground space}. Also, it is clear from expression (\ref{generalizeddeg}) and (\ref{basicdima}), (\ref{generalizeddima}) that $deg. \leq dim.$ ($\partial B\subseteq B$ so that $L\leq N_{b}$). Usually, $deg.$ is much smaller than $dim.$ because the former involves only contributions from boundary vertices of the block while the latter also involves contributions from all bulk vertices. Then as a corollary of the \textbf{Theorem}, we have $D\leq deg.\leq dim.$  

If the block $B$ consists of only one vertex with a spin-$S$, then we conjecture that it is in the maximally entangled state. The \textit{support} has dimension $D=2S+1$.

\section{Examples of the Density Matrix and Open Problems}
\label{secexamples}

The density matrix of the block has been studied in \cite{FKR}, \cite{KHH} and diagonalized directly in \cite{XKHK}, \cite{XKHK2} for $1$-dimensional models, which illustrates the \textbf{Theorem} explicitly. It was shown for different $1$-dimensional AKLT models that the inequality $D\leq deg.$ is always saturated, i.e. $D=deg.$, so that the \textit{support} is exactly equal to the \textit{ground space}. The density matrix is proportional to the projector on the degenerate \textit{ground space} of the block Hamiltonian. Therefore the projector $\mathbf{P}_{D}$ on the \textit{support} of $\rho_{b}$ is equal to the projector $\mathbf{P}_{deg.}$ on the \textit{ground space} of $H_{b}$:
\begin{eqnarray}
	\mathbf{P}_{D}=\mathbf{P}_{deg.} \label{projectors}
\end{eqnarray}
If we denote the identity of the Hilbert space of the block by $\mathbf{I}_{dim.}$, then we also have
\begin{eqnarray}
	\rho_{b}(\mathbf{I}_{dim.}-\mathbf{P}_{deg.})=0. \label{projzero}
\end{eqnarray}
Using these relations (\ref{projectors}) and (\ref{projzero}), the density matrix can be put in the following matrix form
\begin{eqnarray}
	\rho_{b}=\mathbf{\Lambda}\cdot \mathbf{P}_{deg.}=\mathbf{P}_{deg.}\cdot \mathbf{\Lambda}  \label{matrixform}
\end{eqnarray}
where $\mathbf{\Lambda}$ is a diagonal matrix with non-zero eigenvalues of $\rho_{b}$ as entries. It was also proved in \cite{XKHK}, \cite{XKHK2} that in the large block limit $N_{b}\rightarrow\infty$, all eigenvalues become the same so that
\begin{eqnarray}
	\lim_{N_{b}\rightarrow\infty}\mathbf{\Lambda}=\frac{1}{D}\mathbf{I}_{D}, \label{limitlambda}
\end{eqnarray}
where $\mathbf{I}_{D}$ is the identity of the \textit{support}. As a consequence, the density matrix approaches the following limit
\begin{eqnarray}
	\lim_{N_{b}\rightarrow\infty}\rho_{b}\equiv\rho_{\infty}=\frac{1}{deg.}\mathbf{P}_{deg.}=\frac{1}{D}\mathbf{P}_{D}, \label{limit}
\end{eqnarray}
where $\rho_{\infty}$ behaves as the identity in the \textit{ground space} or \textit{support}.

In below we give explicitly the forms of the density matrix obtained in \cite{XKHK}, \cite{XKHK2} and formulate some open problems.

\subsection{$1$-dimensional Basic Model}
\label{secbasicexample}

For the basic model in $1$-dimension, we have spin-$1$'s in the bulk and spin-$\frac{1}{2}$'s at both ends of the chain. The block $B$ consists of $N_{b}$ contiguous bulk vertices and the block Hamiltonian is
\begin{eqnarray}
	H_{b}=\sum^{(N_{b}-1)\ \mbox{\scriptsize{terms}}}_{(l,\ l+1)\ \in\ \mbox{\scriptsize{block}}}\left(\frac{1}{6}\left(\bi{S}_{l}\cdot\bi{S}_{l+1}\right)^{2}+\frac{1}{2} \left(\bi{S}_{l}\cdot\bi{S}_{l+1}\right)+\frac{1}{3}\right). \label{basich1d}
\end{eqnarray}

As shown in \cite{XKHK}, the \textit{ground space} of $H_{b}$ is $4$-dimensional. It can be spanned by $\{|\mbox{G}; \alpha\rangle, \alpha=0,1,2,3\}$, and these are also eigenstates of $\rho_{b}$ with non-zero eigenvalues (see \cite{XKHK} for an explicit construction of these states). In the large block limit, non-zero eigenvalues of the density matrix $\rho_{b}$ become the same and the density matrix is proportional to a $4$-dimensional projector 
\begin{eqnarray}
\lim_{N_{b}\rightarrow\infty}\rho_{b}\equiv\rho_{\infty}=\frac{1}{4}\mathbf{P}_{4}. \label{basiclimit}
\end{eqnarray}
Here $\mathbf{P}_{4}$ is the projector onto the $4$-dimensional \textit{ground space}. Both the von Neumann entropy and Renyi entropy are equal to $\ln4$ in the limit.

For finite block, $\rho_{b}|\mbox{G}; \alpha\rangle=\Lambda_{\alpha}|\mbox{G}; \alpha\rangle$, $\alpha=0,1,2,3$. The eigenvalues of the density matrix are \cite{FKR}, \cite{KHH}, \cite{XKHK}
\begin{eqnarray}
	\Lambda_{\alpha}=\left\{ \begin{array}{cc}
         \frac{1}{4}(1+3(-\frac{1}{3})^{N_{b}}), & \alpha=0;\\ \\
         \frac{1}{4}(1-(-\frac{1}{3})^{N_{b}}), & \alpha=1, 2, 3. \end{array}\right. \label{basiceigenvalue}
\end{eqnarray}
Here $N_{b}$ is the size of the block. The matrix $\mathbf{\Lambda}$ in (\ref{matrixform}) becomes
\begin{eqnarray}
	\mathbf{\Lambda}=\left(\begin{array}{cccc}
         \Lambda_{0} & 0 & 0 & 0 \\
         0 & \Lambda_{1} & 0 & 0 \\
         0 & 0 & \Lambda_{1} & 0 \\
         0 & 0 & 0 & \Lambda_{1}
         \end{array}\right).
\end{eqnarray}
So that the density matrix $\rho_{b}$ is the projector $\mathbf{P}_{4}$ multiplied by this matrix $\mathbf{\Lambda}$.

\subsection{$1$-dimensional Homogeneous Model with Generic Spin}
\label{sechomogeneousexample}

For the high spin homogeneous model in $1$-dimension, we have spin-$S$ in the bulk and spin-$\frac{S}{2}$ at both ends to ensure uniqueness of the ground state. The block Hamiltonian is
\begin{eqnarray}
	H_{b}=\sum^{(N_{b}-1)\ \mbox{\scriptsize{terms}}}_{(l,\ l+1)\ \in\ \mbox{\scriptsize{block}}}\left(\sum^{2S}_{J=S+1} A_{J}(l,l+1)\pi_{J}(l, l+1)\right). \label{homohs}
\end{eqnarray}
Here projector $\pi_{J}(l, l+1)$ is defined in the same way as in (\ref{projectorpi}). Only nearest neighbor $l$ and $(l+1)$ interact, and $A_{J}(l,l+1)$'s are positive coefficients.

As shown in \cite{XKHK}, the \textit{ground space} of $H_{b}$ is $(S+1)^{2}$-dimensional and can be spanned by $\{|\mbox{VBS}_{N_{b}}(J, M)\rangle,\ J=0,1,\ldots,S,\ M=-J,-J+1,\ldots,J\}$. These are also eigenstates of $\rho_{b}$ with non-zero eigenvalues (see \cite{XKHK} for an explicit construction of these \textit{degenerate VBS states} of the block). In the large block limit, all non-zero eigenvalues approach the same value and we have
\begin{eqnarray}
\lim_{N_{b}\rightarrow\infty}\rho_{b}\equiv\rho_{\infty}=\frac{1}{(S+1)^{2}}\mathbf{P}_{(S+1)^{2}}. \label{homoslimit}
\end{eqnarray}
Here $\mathbf{P}_{(S+1)^{2}}$ is the projector onto the degenerate \textit{ground space} of the block Hamiltonian. Both the von Neumann entropy and Renyi entropy are saturated and equal to $\ln[(S+1)^{2}]$ in the limit.

For finite block, $\rho_{b}|\mbox{VBS}_{N_{b}}(J, M)\rangle=\Lambda(J)|\mbox{VBS}_{N_{b}}(J, M)\rangle$ with eigenvalues of the density matrix $\Lambda(J)$ independent of $M$ and given by
\begin{eqnarray}
	\fl \Lambda(J)=\frac{1}{(S+1)^{2}}\left\{1+\sum^{S}_{l=1}(2l+1)\lambda^{N_{b}-1}(l,S)
I_{l}\left(\frac{1}{2}J(J+1)-\frac{1}{2}S(\frac{1}{2}S+1)\right)\right\}. \label{eigenvalues}
\end{eqnarray}
in which $I_{l}$ is an $l^{th}$ order  polynomial given by a recurrence relation \cite{FM}, \cite{KHH}, \cite{XKHK}, and $\lambda(l,S)$ is given by
\begin{eqnarray}
	\lambda(l,S)=\frac{(-1)^l S!(S+1)!}{(S-l)!(S+l+1)!}. \label{lambdals}
\end{eqnarray}
So that matrix $\mathbf{\Lambda}$ consists of these eigenvalues (\ref{eigenvalues}) in diagonal form and the density matrix $\rho_{b}$ is the projector $\mathbf{P}_{(S+1)^{2}}$ multiplied by this matrix $\mathbf{\Lambda}$.

\subsection{$1$-dimensional Generalized (Inhomogeneous) Model}
\label{secgeneraliedexample}

For the generalized model in $1$-dimension, we label the left ending site of the block by $l=1$ with spin value $S_{1}=\frac{1}{2}(M_{01}+M_{12})$ and the right ending site by $l=L$ with spin value $S_{L}=\frac{1}{2}(M_{L-1,L}+M_{L,L+1})$. The block Hamiltonian is
\begin{eqnarray}
	H_{b}=\sum^{(N_{b}-1)\ \mbox{\scriptsize{terms}}}_{(l,\ l+1)\ \in\ \mbox{\scriptsize{block}}}\ \left(\sum^{S_{l}+S_{l+1}}_{J=S_{l}+S_{l+1}-M_{l,l+1}+1} A_{J}(l,l+1)\pi_{J}(l, l+1)\right) \label{generalizedh1d}
\end{eqnarray}
with $\pi_{J}(l,l+1)$ defined in (\ref{projectorpi}) and $A_{J}(l,l+1)$ positive coefficients.

As shown in \cite{XKHK2}, the \textit{ground space} of $H_{b}$ is $(M_{01}+1)(M_{L,L+1}+1)$-dimensional and can be spanned by $\{|\mbox{VBS}_{N_{b}}(J, M)\rangle,\ J=|J_{-}|,|J_{-}|+1,\ldots,J_{+},\ M=-J,-J+1,\ldots,J\}$ (see \cite{XKHK2} for an explicit construction of these \textit{degenerate VBS states} of the block). Here $J_{-}=\frac{1}{2}(M_{01}-M_{L,L+1})$ and $J_{+}=\frac{1}{2}(M_{01}+M_{L,L+1})$. These states are eigenstates of $\rho_{b}$ with non-zero eigenvalues.

In the large block limit, assuming that $M_{01}\rightarrow S_{-}$ and $M_{L,L+1}\rightarrow S_{+}$, we have
\begin{eqnarray}
\lim_{N_{b}\rightarrow\infty}\rho_{b}\equiv\rho_{\infty}=\frac{1}{(S_{-}+1)(S_{+}+1)}\mathbf{P}_{(S_{-}+1)(S_{+}+1)}. \label{generalizedlimit}
\end{eqnarray}
Here $S_{-}$ and $S_{+}$ are the first and last spins in the block, respectively. The von Neumann entropy is equal to the Renyi entropy
\begin{eqnarray}
	S_{v.N}=S_{R}=\ln[(S_{-}+1)(S_{+}+1)] \label{limitentropy}
\end{eqnarray}
in the limit.

For finite block, $\rho_{b}|\mbox{VBS}_{N_{b}}(J, M)\rangle=\Lambda(J)|\mbox{VBS}_{N_{b}}(J, M)\rangle$.  The eigenvalues $\Lambda(J)$ are independent of $M$ and given by
\begin{eqnarray}
	\fl \Lambda(J)=
\frac{1}{(M_{01}+1)(M_{L,L+1}+1)}\left\{1+\sum^{M_{<}}_{l=1}(2l+1)\left[\prod^{N_{b}-1}_{j=1}\lambda(l,M_{j,j+1})\right]Poly.(l,J)\right\}, \label{expansion}
\end{eqnarray}
in which $M_{<}=Min\{M_{j,j+1},\ j=1,\ldots,L-1\}$ being the minimum of the multiplicity numbers, $Poly.(l,J)$ is a polynomial of $J$ depending on $l$, and $\lambda(l,M_{j,j+1})$ is given by
\begin{eqnarray}
	\lambda(l,M_{ij})=\frac{(-1)^l M_{ij}!(M_{ij}+1)!}{(M_{ij}-l)!(M_{ij}+l+1)!}. \label{lambdalm}
\end{eqnarray}
So that matrix $\mathbf{\Lambda}$ consists of these eigenvalues (\ref{expansion}) in diagonal form. The density matrix is the projector $\mathbf{P}_{(S_{-}+1)(S_{+}+1)}$ multiplied by the matrix $\mathbf{\Lambda}$. 

\subsection{Some Open Problems}
\label{secproblems}

One open problem is the calculation of non-zero eigenvalues of the density matrix $\rho_{b}$ for general and more complicated graphs. One should start with the Cayley tree (also known as the Bethe tree). We expect that an exact explicit expression for the non-zero eigenvalues is possible, because it has no loops. It is also important to calculate non-zero eigenvalues of $\rho_{b}$ for graphs with loops. We expect that in the large block limit, each non-zero eigenvalue should approach the same value $\frac{1}{D}=\frac{1}{deg.}$ and the entanglement entropies should be saturated, i.e. $S_{v.N}=S_{R}=\ln{D}=\ln{(deg.)}$.

\section{Conclusion}
\label{sceconclusion}

We have studied the entanglement of the AKLT model. We formulate the AKLT model on an arbitrary connected graph. The Hamiltonian (\ref{basich}), (\ref{generalizedh}) is a sum of projectors which describe interactions between nearest neighbors. The condition of uniqueness of the ground state relates the spin value at each vertex with multiplicity numbers associated with edges incident (connected) to the vertex, see (\ref{basiccond}), (\ref{condition1}), (\ref{condition2}). The unique ground state is known as the Valence-Bond-Solid state (\ref{basicvbs}), (\ref{generalizedvbs}). The VBS state can be used instead of the cluster state in measurement-based quantum computation, see \cite{BM}.

To study the entanglement, the graph is divided into two parts: the \textit{block} and the \textit{environment}. We investigate the density matrix $\rho_{b}$ of the block and show that it has many zero eigenvalues. We describe the subspace (called the \textit{support}) of eigenvectors of $\rho_{b}$ with non-zero eigenvalues. We have proved (see \textbf{Theorem} in Section \ref{secmatrix}) that this subspace is the degenerate \textit{ground space} of some Hamiltonian, we call it the \textit{block} Hamiltonian (\ref{subsystemh}).

The entanglement can be measured by the von Neumann entropy or the Renyi entropy of the density matrix $\rho_{b}$. Most eigenvalues of $\rho_{b}$ vanish and have no contribution to the entanglement entropies. The density matrix takes the form of a projector on the \textit{ground space} multiplied by another matrix (conjectured in \cite{XKHK} for an arbitrary graph).

Non-zero eigenvalues of $\rho_{b}$ have been calculated for a variety of $1$-dimensional AKLT models \cite{FKR}, \cite{KHH}, \cite{XKHK}, \cite{XKHK2}. They are given as illustrations. We find that in these cases the \textit{support} coincide with the \textit{ground space}, so their dimensions are equal $D=deg.$ In the large block limit, all non-zero eigenvalues become the same and the density matrix is proportional to a projector (\ref{basiclimit}), (\ref{homoslimit}), (\ref{generalizedlimit}). The von Neumann entropy equals the Renyi entropy and both take the saturated value $S_{v.N}=S_{R}=\ln D=\ln (deg.)$.

For more complicated graphs, non-zero eigenvalues of the density matrix are still unknown. One open problem is to calculate those eigenvalues. One may start with the Cayley tree because there is no loops and we expect to obtain exact explicit expressions of the eigenvalues.

\ack

The authors would like to thank S. Bravyi, H. Fan, T. Hirano, H. Katsura, A. N. Kirillov and F. Verstraete for valuable discussions and suggestions. The work is supported by NSF Grant DMS-0503712.

\appendix

\section{Coherent State Basis}
\label{secapp}

Given the VBS state in the Schwinger representation (\ref{generalizedvbs}), it is possible to use a coherent state basis \cite{AAH}, \cite{KK}, \cite{FM}, \cite{KHH}, \cite{XKHK}. Then spin operators behave like classical unit vectors. The coherent state basis may simplify the calculation of correlation functions, eigenvalues and eigenstates of the density matrix.

We introduce spinor coordinates
\begin{eqnarray}
	\left(u, v\right)=\left(\cos\frac{\theta}{2}\rme^{\rmi\frac{\phi}{2}},\ \sin\frac{\theta}{2}\rme^{-\rmi\frac{\phi}{2}}\right), \qquad 0\leq\theta\leq\pi, \quad 0\leq\phi\leq 2\pi. \label{spinor}
\end{eqnarray}
Then for a point $\hat{\Omega}\equiv (\sin\theta\cos\phi, \sin\theta\sin\phi, \cos\theta)$ on the unit sphere, the spin-$S$ coherent state is defined as
\begin{eqnarray}
	|\hat{\Omega}\rangle=\frac{\left(ua^{\dagger}+vb^{\dagger}\right)^{2S}}{\sqrt{\left(2S\right)!}}|\mbox{vac}\rangle. \label{coherent}
\end{eqnarray}
Here we have fixed the overall phase (a $U(1)$ gauge degree of freedom) since it has no physical content. Note that (\ref{coherent}) is covariant under $SU(2)$ transforms (see \cite{H}, \cite{XKHK}). The set of coherent states is complete (but not orthogonal) such that \cite{ACGT}, \cite{FM}
\begin{eqnarray}
	\frac{2S+1}{4\pi}\int \rmd\hat{\Omega} |\hat{\Omega}\rangle \langle \hat{\Omega}|=\sum^{S}_{m=-S}|S, m\rangle\langle S, m|=I_{2S+1}, \label{completeness}
\end{eqnarray}
where $|S, m\rangle$ denote the eigenstate of $\bi{S}^{2}$ and $S_{z}$, and $I_{2S+1}$ is the identity of the $(2S+1)$-dimensional Hilbert space for spin-$S$. The completeness relation (\ref{completeness}) can be used in taking trace of an arbitrary operator.

Take the partition function $\Phi=\Tr[\ |\mbox{VBS}\rangle\langle\mbox{VBS}|\ ]$ for example. Using the coherent state basis (\ref{coherent}) and realizing that
\begin{eqnarray}
	\langle \mbox{vac}|a^{S+m}b^{S-m}|\hat{\Omega}\rangle=\sqrt{(2S)!}\ u^{S+m}v^{S-m}, \label{innerprod}
\end{eqnarray}
we have
\begin{eqnarray}
	\Phi &=&\Tr[\ |\mbox{VBS}\rangle\langle \mbox{VBS}|\ ] \nonumber \\
	&=&\left[\prod_{l}\frac{(2S_{l}+1)!}{4\pi}\right]\int \left[\prod_{l}\rmd\hat{\Omega}_{l}\right]\prod_{\langle kl\rangle}\left[\frac{1}{2}\left(1-\hat{\Omega}_{k}\cdot\hat{\Omega}_{l}\right)^{M_{kl}}\right]. \label{partition}
\end{eqnarray}
Here $\langle kl\rangle$ runs over all edges and we are integrating the classical solid angle
$\hat{\Omega}_{l}$ over each vertex $l$.

\section{Ground State Degeneracy of the Block Hamiltonian}
\label{secapp2}

We prove Katsura's formula (\ref{generalizeddeg}) for the degeneracy of ground states of the block Hamiltonian. Our proof is based on \cite{Kcomm}. The block Hamiltonian is defined in (\ref{subsystemh}). We first look at the uniqueness condition (\ref{condition1}). For an arbitrary vertex $l$ in the block $B$, the condition can be written as
\begin{eqnarray}
	2S_{l}=\sum_{k}M_{kl}=\sum_{k\in B}M_{kl}+\sum_{k\in \partial E}M_{kl}, \qquad l\in B. \label{split}
\end{eqnarray}
Note that the sum over vertices $k\in \partial E$ is outside the block $B$. These terms are only present for boundary vertices $l\in \partial B$. Expression (\ref{split}) is valid for \textit{any} vertex in the block (for a bulk vertex the last summation vanishes). Next we define \textit{the block VBS state}
\begin{eqnarray}
	|\mbox{VBS}_{N_{b}}\rangle=\prod_{\langle kl\rangle\in B}
\left(a^{\dagger}_{k}b^{\dagger}_{l}-b^{\dagger}_{k}a^{\dagger}_{l}\right)^{M_{kl}}|\mbox{vac}\rangle, \qquad k\in B,\quad l\in B. \label{blockvbs}
\end{eqnarray}
Here edge $\langle kl\rangle$ lies completely inside the block $B$. Now an \textit{arbitrary} ground state of the block Hamiltonian $H_{b}$ takes the following form
\begin{eqnarray}
	|\mbox{G}\rangle=\left[\prod^{L\ \mbox{\scriptsize{terms}}}_{l\in\ \partial B}f(a^{\dagger}_{l},b^{\dagger}_{l})\right]|\mbox{VBS}_{N_{b}}\rangle, \label{ground}
\end{eqnarray}
where $f(a^{\dagger}_{l},b^{\dagger}_{l})$ is a polynomial (may depend on the vertex $l$) in $a^{\dagger}_{l}$ and $b^{\dagger}_{l}$ and the product runs over all boundary vertices (with the number denoted by $L$). The degree of this polynomial is equal to $\sum_{k\in \partial E}M_{kl}$. (Each term in the polynomial must have the same total power $\sum_{k\in \partial E}M_{kl}$ of $a^{\dagger}_{l}$ and $b^{\dagger}_{l}$.) It is straightforward to verify that $|\mbox{G}\rangle$ in (\ref{ground}) is a ground state: 
\begin{enumerate}
	\item The power of $a^{\dagger}_{l}$ and $b^{\dagger}_{l}$ in $|\mbox{VBS}_{N_{b}}\rangle$ is $\sum_{k\in B}M_{kl}$ (see (\ref{blockvbs})) so that the total power of $a^{\dagger}_{l}$ and $b^{\dagger}_{l}$ in (\ref{ground}) is $\sum_{k\in B}M_{kl}+\sum_{k\in \partial E}M_{kl}=2S_{l}$ according to (\ref{split}). Therefore, we have the correct power $2S_{l}$ of the bosonic operators $a^{\dagger}_{l}$ and $b^{\dagger}_{l}$ for each vertex $l$ in the block $B$ (constraint (\ref{constraint}) is satisfied);
	\item There is no projection on any edge spin value greater than $S_{k}+S_{l}-M_{kl}+1$ because of the construction of the block VBS state (\ref{blockvbs}). (One could use the same reasoning as in Section \ref{secgeneralized}).
\end{enumerate}
Therefore the degeneracy $deg.$ of the ground states of $H_{b}$ is equal to the number of linearly independent states of the form (\ref{ground}). Since $a^{\dagger}_{l}$'s and $b^{\dagger}_{l}$'s are bosonic and commute, the number of linearly independent polynomials $f(a^{\dagger}_{l},a^{\dagger}_{l})$ for an arbitrary $l$ is equal to its degree plus one, i.e. $\left(\sum_{k\in \partial E}M_{kl}\right)+1$, $\forall\ l\in \partial B$. So that the total number of linearly independent polynomials of the form $\prod^{L\ \mbox{\scriptsize{terms}}}_{l\in\ \partial B}f(a^{\dagger}_{l},b^{\dagger}_{l})$ is the product of these numbers for each $l\in \partial B$. Finally, the ground state degeneracy of the block Hamiltonian $H_{b}$ is (Katsura's formula \cite{Kcomm})
\begin{eqnarray}
	deg.=\prod_{l\in\partial B}\left[\left(\sum_{k\in \partial E}M_{kl}\right)+1\right]. \label{kastura}
\end{eqnarray}
In the case of the basic model all $M_{kl}=1$, formula (\ref{kastura}) has a graphical illustration, see Figure (\ref{figure2}). We count the number \# of all cut edges (dashed lines) incident to one boundary vertex of the block, then add one to the number \#. The degeneracy is the product of these $(\# +1)$'s for each boundary vertex.



\section*{References}

\begin{thebibliography}{99}

\bibitem{AKLT1}
Affleck A, Kennedy T, Lieb E H and Tasaki H 1987 Rigorous Results on Valence-Bond Ground States in Antiferromagnets \textit{Phys. Rev. Lett.} \textbf{59} 799-802

\bibitem{AKLT2}
Affleck A, Kennedy T, Lieb E H and Tasaki H 1988 Valence Bond Ground States in Isotropic Quantum Antiferromagnets \textit{Commun. Math. Phys.} \textbf{115} 477-528

\bibitem{AFOV}
Amico L, Fazio R, Osterloh A and Vedral V 2008 Entanglement in many-body systems \textit{Rev. Mod. Phys.} \textbf{80} 517-576

\bibitem{ACGT}
Arecchi F T, Courtens E, Gilmore R and Thomas H 1972 Atomic Coherent States in Quantum Optics \textit{Phys. Rev.} A\,\textbf{6} 2211-2237

\bibitem{AAH}
Arovas D P, Auerbach A and Haldane F D M 1988 Extended Heisenberg models of antiferromagnetism: Analogies to the fractional quantum Hall effect \textit{Phys. Rev. Lett.} \textbf{60} 531-534

\bibitem{AEPW}
Audenaert K, Eisert J, Plenio M B and Werner R F 2002 Entanglement Properties of the Harmonic Chain \textit{Phys. Rev.} A\,\textbf{66} 042327 \textit{Preprint} arXiv:quant-ph/0205025.

\bibitem{A}
Auerbach A 1998 \textit{Interacting Electrons and Quantum Magnetism} (New York: Springer)

\bibitem{BDV}
Barthel T, Dusuel S and Vidal J 2006 Entanglement Entropy beyond the Free Case \textit{Phys. Rev. Lett.} \textbf{97} 220402

\bibitem{BD}
Bennett C H and DiVincenzo D P 2000 Quantum information and computation \textit{Nature} \textbf{404} 247-255

\bibitem{BM}
Brennen G K and Miyake A 2008 Measurement-based quantum computer in the gapped ground state of a two-body Hamiltonian \textit{Preprint} arXiv:0803.1478

\bibitem{CDR}
Campos Venuti L, Degli Esposti Boschi C and Roncaglia M 2006 Long-distance entanglement in spin systems \textit{Phys. Rev. Lett.} \textbf{96} 247206

\bibitem{CTZ}
Cappelli A, Trugenberger C A and Zemba G R 1993 Infinite Symmetry in the Quantum Hall Effect 	\textit{Nucl. Phys.} B\, \textbf{396} 465-490 \textit{Preprint}	arXiv:hep-th/9206027v1

\bibitem{CCK}
Chayes J, Chayes L and Kivelson S 1989 Valence bond ground states in a frustrated two-dimensional spin 1/2 Heisenberg antiferromagnet \textit{Commun. Math. Phys.} \textbf{123} 53-83

\bibitem{CZWZ}
Chen Y, Zanardi P, Wang Z D and Zhang F C 2004 Entanglement and Quantum Phase Transition in Low Dimensional Spin Systems \textit{New Journal of Physics} \textbf{8} 97 \textit{Preprint} arXiv:quant-ph/0407228; Zhao Y, Zanardi P and Chen G 2004 Quantum Entanglement and the Self-Trapping Transition in Polaronic Systems \textit{Preprint} arXiv:quant-ph/0407080; Hamma A, Ionicioiu R and Zanardi P 2005 Ground state entanglement and geometric entropy in the Kitaev's model \textit{Phys.Lett.} A\,\textbf{337} 22 \textit{Preprint} arXiv:quant-ph/0406202; Giorda P and Zanardi P 2003 Ground-State Entanglement in Interacting Bosonic Graphs \textit{Preprint} arXiv:quant-ph/0311058

\bibitem{CEP}
Cramer M, Eisert J and Plenio M B 2007 Statistical Dependence of Entanglement-Area Laws \textit{Phys. Rev. Lett.} \textbf{98} 220603 \textit{Preprint} arXiv:quant-ph/0611264

\bibitem{CEPD}
Cramer M, Eisert J, Plenio M B and Dreisig J 2006 On an entanglement-area-law for general bosonic harmonic lattice systems \textit{Phys. Rev.} A\,\textbf{73} 012309 \textit{Preprint} arXiv:quant-ph/0505092

\bibitem{FK}
Fan H and Korepin V E 2008 Quantum Entanglement of the Higher Spin Matrix Product States (in preparation)

\bibitem{FL}
Fan H and Lloyd S 2005 Entanglement of eta-pairing state with off-diagonal long-range order \textit{J.Phys.} A\,\textbf{38} 5285 \textit{Preprint} arXiv:quant-ph/0405130

\bibitem{FKR}
Fan H, Korepin V E and Roychowdhury V 2004 Entanglement in a Valence-Bond-Solid State \textit{Phys. Rev. Lett.} \textbf{93} 227203 \textit{Preprint} arXiv:quant-ph/0406067

\bibitem{FKR2}
Fan H, Korepin V E, and Roychowdhury V 2005 Valence-Bond-Solid state entanglement in a 2-D Cayley tree \textit{Preprint} arXiv:quant-ph/0511150

\bibitem{FKRHB}
Fan H, Korepin V E, Roychowdhury V, Hadley C and Bose S 2007 Boundary effects to the entanglement entropy and two-site entanglement of the spin-1 valence-bond solid \textit{Phys. Rev.} B\,\textbf{76} 014428 \textit{Preprint} arXiv:quant-ph/0605133

\bibitem{FNW0}
Fannes M, Nachtergaele B and Werner R F, 1989 Exact Antiferromagnetic Ground
States for Quantum Chains \textit{Europhys. Lett.} \textbf{10}, 633-637

\bibitem{FNW1}
Fannes M, Nachtergaele B and Werner R F 1991 Valence Bond States on Quantum
Spin Chains as Ground States with Spectral Gap \textit{J. Phys.} A\, \textbf{24}
L185-L190

\bibitem{FNW}
Fannes M, Nachtergaele B and Werner R F 1992 Finitely correlated states on quantum spin chains \textit{Commun. Math. Phys.} \textbf{144} 443-490

\bibitem{FNW2}
Fannes M, Nachtergaele B and Werner R F, 1992 Ground States of VBS models on
Cayley Trees \textit{J. Stat. Phys.} \textbf{66} 939-973

\bibitem{FIJK}
Franchini F, Its A R, Jin B-Q and Korepin V E 2007 Ellipses of Constant Entropy in the XY Spin Chain \textit{J. Phys.} A\,\textbf{40} 8467 \textit{Preprint} arXiv:quant-ph/0609098v5

\bibitem{FIK}
Franchini F, Its A R and Korepin V E 2008 Renyi Entropy of the XY Spin Chain \textit{J. Phys.} A\,\textbf{41} 025302 \textit{Preprint} arXiv:0707.2534

\bibitem{FM}
Freitag W-D and Muller-Hartmann E 1991 Complete analysis of two spin correlations of valence bond solid chains for all integer spins \textit{Z. Phys.} B\,\textbf{83} 381

\bibitem{GMC}
Garcia-Ripoll J J, Martin-Delgado M A and Cirac J I 2004 Implementation of Spin Hamiltonians in Optical Lattices \textit{Phys. Rev. Lett.} \textbf{93} 250405 \textit{Preprint} arXiv:cond-mat/0404566

\bibitem{GRAC}
Ghosh S, Rosenbaum T F, Aeppli G and Coppersmith S N 2003 Entangled quantum state of magnetic dipoles \textit{Nature} \textbf{425} 48-51

\bibitem{GLL}
Gu S-J, Lin H-Q and Li Y-Q 2003 Entanglement, quantum phase transition, and scaling in the XXZ chain \textit{Phys. Rev.} A\,\textbf{68} 042330

\bibitem{GDLL}
Gu S-J, Deng S-S, Li Y-Q and Lin H-Q 2004 Entanglement and Quantum Phase Transition in the Extended Hubbard Model \textit{Phys. Rev. Lett.} \textbf{93} 086402

\bibitem{Had}
Hadley C 2008 Single copy entanglement in a gapped quantum spin chain \textit{Phys. Rev. Lett.} \textbf{100} 170001 \textit{Preprint} arXiv:0801.3681

\bibitem{HKAHB}
Hagiwara M, Katsumata K, Affleck I, Halperin B I and Renard J P 1990
Observation of S=1/2 degrees of freedom in an S=1 linear-chain Heisenberg antiferromagnet 
\textit{Phys. Rev. Lett.} \textbf{65} 3181

\bibitem{H}
Haldane F D M 1983 Continuum dynamics of the 1-D Heisenberg antiferromagnet: Identification with the $O(3)$ nonlinear sigma model \textit{Phys. Lett.} \textbf{93A} 464-468; 1983 Nonlinear Field Theory of Large-Spin Heisenberg Antiferromagnets: Semiclassically Quantized Solitons of the One-Dimensional Easy-Axis N¨¦el Stat e\textit{Phys. Rev. Lett.} \textbf{50} 1153-1156

\bibitem{HR0}
Haldane F D M and Rezayi E H 1985 Finite-Size Studies of the Incompressible State of the Fractionally Quantized Hall Effect and its Excitations \textit{Phys. Rev. Lett.} \textbf{54} 237

\bibitem{HR}
Haldane F D M and Rezayi E H 1988 Spin-Singlet Wave Function for the Half-Integral Quantum Hall Effect \textit{Phys. Rev. Lett.} \textbf{60} 956

\bibitem{Ha}
Hamermesh M 1989 \textit{Group Theory and Its Application to Physical Problems} (New York: Dover Publications)

\bibitem{HZS}
Haque M, Zozulya O and Schoutens K 2007  Entanglement Entropy in Fermionic Laughlin States \textit{Phys. Rev. Lett.} \textbf{98} 060401 \textit{Preprint}	arXiv:cond-mat/0609263v2

\bibitem{Har}
Harary F 1969 \textit{Graph Theory} (Addison-Weslay Publ. Comp. Reading Massachusets)

\bibitem{Has}
Hastings M B 2007 An Area Law for One Dimensional Quantum Systems \textit{J. Stat. Mech.} P08024

\bibitem{H2}
Hirano T and Hatsugai Y 2007 Entanglement Entropy of One-dimensional Gapped Spin Chains \textit{J. Phys. Soc. Jpn.} \textbf{76} 074603

\bibitem{HLW}
Holzhey C, Larsen F and Wilczek F 1994 Geometric and Renormalized Entropy in Conformal Field Theory \textit{Nucl. Phys.} B\,\textbf{424} 443 \textit{Preprint} arXiv:hep-th/9403108v1

\bibitem{ILO}
Iblisdir S, Latorre J I and Orus R 2007 Entropy and Exact Matrix Product Representation of the Laughlin Wave Function \textit{Phys.Rev.Lett.} \textbf{98} 060402 \textit{Preprint} cond-mat/0609088

\bibitem{IT}
Iske P L and Caspers W J 1987 Exact ground states of one-dimensional valence-bond-solid
Hamiltonians \textit{Mod. Phys. Lett.} \textbf{B1} 231-237

\bibitem{IJK}
Its  A R, Jin B-Q and Korepin V E 2005 Entanglement in XY Spin Chain \textit{J. Phys.} A\,\textbf{38} 2975 \textit{Preprint} arXiv:quant-ph/0409027

\bibitem{JK}
Jin B-Q and Korepin V E 2004 Quantum Spin Chain, Toeplitz Determinants and Fisher-Hartwig Conjecture. \textit{J. Stat. Phys.} \textbf{116} 79 \textit{Preprint} arXiv:quant-ph/0304108

\bibitem{Kcomm}
Katsura H 2008 Private Communications (not published).

\bibitem{KHH}
Katsura H, Hirano T and Hatsugai Y 2007 Exact Analysis of Entanglement in Gapped Quantum Spin Chains \textit{Phys. Rev.} B\,\textbf{76} 012401 \textit{Preprint} arXiv:cond-mat/0702196

\bibitem{KHK}
Katsura H, Hirano T and Korepin V E 2007 Entanglement in an SU(n) Valence-Bond-Solid State \textit{Preprint} arXiv:0711.3882

\bibitem{KM}
Keating J P and Mezzadri F 2004 Random Matrix Theory and Entanglement in Quantum Spin Chains \textit{Commun. Math. Phys.} \textbf{252} 543-579 \textit{Preprint} arXiv:quant-ph/0407047

\bibitem{KLT}
Kennedy T, Lieb E H and Tasaki H 1988 A two-dimensional isotropic quantum antiferromagnet with unique disordered ground state \textit{J. Stat. Phys.} \textbf{53} 383-415

\bibitem{KK}
Kirillov A N and Korepin V E 1990 Correlation Functions in Valence Bond Solid Ground State \textit{Sankt Petersburg Mathematical Journal} \textbf{1} 47; 1989 \textit{Algebra and Analysis} \textbf{1} 47

\bibitem{KP}
Kitaev A and Preskill J 2006 Topological Entanglement Entropy \textit{Phys. Rev. Lett.} \textbf{96} 110404

\bibitem{K}
Korepin V E 2004 Universality of Entropy Scaling in One Dimensional Gapless Models \textit{Phys. Rev. Lett.} \textbf{92} 096402 \textit{Preprint} arXiv:cond-mat/0311056

\bibitem{LO}
Latorre J I and Orus R 2004 Adiabatic quantum computation and quantum phase transitions
\textit{Phys. Rev.} A\,\textbf{69} 062302 \textit{Preprint} quant-ph/0308042.

\bibitem{LORV}
Latorre J I, Orus R, Rico E and Vidal J 2005 Entanglement entropy in the Lipkin-Meshkov-Glick model \textit{Phys. Rev.} A\,\textbf{71} 064101 \textit{Preprint} cond-mat/0409611

\bibitem{LRV}
Latorre J I, Rico E and Vidal G 2004 Ground state entanglement in quantum spin chains \textit{Quant. Inf. Comp.} \textbf{4} 048

\bibitem{LA}
Laughlin R B 1983 Anomalous Quantum Hall Effect: An Incompressible Quantum Fluid with Fractionally Charged Excitations \textit{Phys. Rev. Lett.} \textbf{50} 1395

\bibitem{LW}
Levin M and Wen X G 2006 Detecting Topological Order in a Ground State Wave Function \textsl{Phys. Rev. Lett.} \textbf{96} 110405

\bibitem{L}
Lloyd S 1993 \textit{Science} A potentially realizable quantum computer \textbf{261} 1569-1671; 1994 Envisioning a Quantum Supercomputer. \textit{ibid}. \textbf{263} 695

\bibitem{MN}
Michalakis S and Nachtergaele B 2006 Entanglement in Finitely Correlated Spin States \textit{Phys. Rev. Lett.} \textbf{97} 140601 \textit{Preprint} arXiv:math-ph/0606018

\bibitem{N1}
Nachtergaele B 1990 \textit{Working with Quantum Markov States and their Classical Analogues}
in Accardi L and Waldenfels W V (Eds.) \textit{Quantum Probability and Applications
V} (LNM 1442) (Springer Verlag: Berlin-Heidelberg-New York)

\bibitem{N2}
Nachtergaele B 1996 The spectral gap for some quantum spin chains with discrete symmetry breaking \textit{Commun. Math. Phys.} \textbf{175} 565-606

\bibitem{NC}
Nielsen M A and Chuang I L 2000 \textit{Quantum Computation and Quantum
Information} (Cambridge: Cambridge Univ. Press)

\bibitem{O1}
Orus R 2005 Entanglement and majorization in (1+1)-dimensional quantum systems \textit{Phys.Rev}. A\,\textbf{71} 052327 \textit{Preprint} quant-ph/0501110 

\bibitem{OL}
Orus R and Latorre J I 2004 Universality of entanglement and quantum-computation complexity \textit{Phys. Rev.} A\,\textbf{69} 052308

\bibitem{ON}
Osborne T J and Nielsen M A 2002 Entanglement in a simple quantum phase transition \textit{Phys. Rev.} A\,\textbf{66} 032110

\bibitem{OAFF}
Osterloh A, Amico L, Falci G and Fazio R 2002 Scaling of entanglement close to quantum phase transitions \textit{Nature} \textbf{416} 608

\bibitem{PP}
Pachos J K and Plenio M B 2004 Three-Spin Interactions in Optical Lattices and Criticality in Cluster Hamiltonians \textit{Phys. Rev. Lett.} \textbf{93} 056402

\bibitem{PEDC}
Plenio M B, Eisert J, Dreisig J and Cramer M 2005 Geometric entropy in harmonic lattice systems \textit{Phys. Rev. Lett.} \textbf{94} 060503

\bibitem{PHE}
Plenio M B, Hartley J and Eisert J 2004 Dynamics and manipulation of entanglement in coupled harmonic systems with many degrees of freedom \textit{New. J. Phys.} \textbf{6} 36 \textit{Preprint} arXiv:quant-ph/0402004

\bibitem{PS}
Popkov V and Salerno M 2004 Logarithmic divergence of the block entanglement entropy for the ferromagnetic Heisenberg model \textit{Preprint} arXiv:quant-ph/0404026

\bibitem{RB}
Rico E and Briegel H J 2007 2D Multipartite Valence Bond States in Quantum Antiferromagnets \textit{Preprint} arXiv:0710.2349

\bibitem{RH}
Ryu S and Hatsugai Y 2006 Entanglement entropy and the Berry phase in the solid state \textit{Phys. Rev.} B\,\textbf{73} 245115

\bibitem{VC}
Verstraete F and Cirac J I 2004 Valence-bond states for quantum computation \textit{Phys. Rev.} A\,\textbf{70} 060302(R) \textit{Preprint} arXiv:quant-ph/0311130v1

\bibitem{VMC}
Verstraete F, Martin-Delgado M A and Cirac J I 2004 Diverging Entanglement Length in Gapped Quantum Spin Systems \textit{Phys. Rev. Lett.} \textbf{92} 087201

\bibitem{VPC}
Verstraete F, Popp M and Cirac J I 2004 Entanglement versus Correlations in Spin Systems \textit{Phys. Rev. Lett.} \textbf{92} 027901; Verstraete F, Porras D and Cirac J I 2004 \textit{ibid}. \textbf{93} 227205

\bibitem{VDB}
Vidal J, Dusuel S and Barthel T 2007 Entanglement entropy in collective models \textit{J. Stat. Mech.} P01015


\bibitem{WK}
Wang J and Kais S 2004 Scaling of entanglement at quantum phase transition for two-dimensional array of quantum dots 2004 arXiv:quant-ph/0405087; 2005 Scaling of entanglement in finite arrays of exchange-coupled quantum dots \textit{Preprint} arXiv:quant-ph/0405085

\bibitem{WKRL}
Wang J, Kais S, Remacle F and Levine R D 2004 Size effects in the electronic properties of finite arrays of exchange coupled quantum dots: A renormalization group approach \textit{Preprint} arXiv:quant-ph/0405088

\bibitem{XKHK}
Xu Y, Katsura H, Hirano T and Korepin V E 2008 Entanglement and Density Matrix of a Block of Spins in AKLT Model \textit{Preprint} arXiv:0802.3221

\bibitem{XKHK2}
Xu Y, Katsura H, Hirano T and Korepin V E 2008 Block Spin Density Matrix of the Inhomogeneous AKLT Model \textit{Preprint} arXiv:0804.1741

\bibitem{ZR}
Zanardi P and Rasetti M 1999 Holonomic Quantum Computation. \textit{Phys. Lett.} A\, \textbf{264} 94-99 \textit{Preprint}	arXiv:quant-ph/9904011v3; Marzuoli A and Rasetti M 2002 Spin network quantum simulator. \textit{Phys. Lett.} A\,\textbf{306} 79-87 \textit{Preprint} arXiv:quant-ph/0209016v1; Rasetti M 2002 A consistent Lie algebraic representation of quantum phase and number operators \textit{Preprint} arXiv:cond-mat/0211081



\end{thebibliography}
\end{document}